\documentclass[twocolumn, times]{aastex62}

\newcommand{\kms}{km s$^{-1}$} 
\newcommand{\psiG}{$\Psi_{\rm GMOS}$}
\newcommand{\psiM}{$\Psi_{\rm Mitchell}$}
\newcommand{\PAkinG}{PA$_{\rm kin}^{\rm GMOS}$}
\newcommand{\PAkinM}{PA$_{\rm kin}^{\rm Mitchell}$}

\graphicspath{{./}{figures/}}

\received{XXXX}
\revised{YYYY}
\accepted{ZZZZ}

\shorttitle{MASSIVE XIV -- GMOS IFS Kinemetry}
\shortauthors{Ene et al.}

\usepackage[utf8x]{inputenc}
\usepackage{graphicx}
\usepackage{amssymb}
\usepackage{amsmath}
\usepackage{natbib}
\begin{document}

\title{The MASSIVE Survey XIV -- Stellar Velocity Profiles and Kinematic Misalignments from 200 pc to 20 kpc in Massive Early-type Galaxies}

\correspondingauthor{Irina Ene}
\email{irina.ene@berkeley.edu}

\author{Irina Ene}
\affiliation{Department of Astronomy, University of California, Berkeley, CA 94720, USA}
\affiliation{Department of Physics, University of California, Berkeley, CA 94720, USA}

\author{Chung-Pei Ma}
\affiliation{Department of Astronomy, University of California, Berkeley, CA 94720, USA}
\affiliation{Department of Physics, University of California, Berkeley, CA 94720, USA}

\author{Jonelle L. Walsh}
\affiliation{George P. and Cynthia Woods Mitchell Institute for Fundamental Physics and Astronomy, and Department of Physics and Astronomy, \\
Texas A\&M University, College Station, TX 77843, USA}

\author{Jenny E. Greene}
\affiliation{Department of Astrophysical Sciences, Princeton University, Princeton, NJ 08544, USA}

\author{Jens Thomas}
\affiliation{Max Plank-Institute for Extraterrestrial Physics, Giessenbachstr. 1, D-85741 Garching, Germany}

\author{John P. Blakeslee}
\affiliation{Gemini Observatory, Casilla 603, La Serena, Chile}

\begin{abstract}

We use high spatial resolution stellar velocity maps from the Gemini GMOS integral-field spectrograph (IFS) and wide-field velocity maps from the McDonald Mitchell IFS to study the stellar velocity profiles and kinematic misalignments from $\sim 200$ pc to $\sim 20$ kpc in 20 early-type galaxies with stellar mass $M_* > 10^{11.7} M_\odot$ in the MASSIVE survey.
While 80\% of the galaxies have low spins ($\lambda < 0.1$) and low rotational velocities ($< 50$ \kms) in both the central region and the main body, we find a diverse range of velocity features and misalignment angles.  For the 18 galaxies with measurable central kinematic axes, 10 have well aligned kinematic axis and photometric major axis, and the other 8 galaxies have misalignment angles that are distributed quite evenly from $15\degr$ to the maximal value of $90\degr$.
There is a strong correlation between central  kinematic misalignment and galaxy spin, where all 4 galaxies with significant spins have well aligned kinematic and photometric axes, but only 43\% of the low-spin galaxies are well aligned.  
The central and main-body kinematic axes within a galaxy are not always aligned.
When the two kinematic axes are aligned ($\sim 60$\% of the cases), they are either also aligned with the photometric major axis or orthogonal to it.
We find 13 galaxies to also exhibit noticeable local kinematic twists, and one galaxy to have a counter-rotating core.
A diverse assembly history consisting of multiple gas-poor mergers of a variety of progenitor mass ratios and orbits
is likely to be needed to account for 
the predominance of low spins and the wide range of central and main-body velocity features reported here
for local massive ETGs.

\end{abstract}


\section{Introduction} 
\label{sec:intro}

Integral field spectroscopic (IFS) surveys have made significant progress in measuring spatially-resolved kinematical properties of local early-type galaxies (ETGs) on typical scales of one effective radius, e.g., SAURON 
\citep{dezeeuwetal2002},
ATLAS$^{\rm 3D}$ \citep{Cappellarietal2011}, SAMI \citep{Croometal2012}, CALIFA \citep{Sanchezetal2012}, MASSIVE \citep{Maetal2014} and MaNGA \citep{Bundyetal2015}. An additional few wide-field IFS or multislit studies of a smaller sample of ETGs reached $\sim 2-4$ effective radii (e.g., \citealt{Brodieetal2014, Raskuttietal2014, Edwardsetal2016}), and a few other IFS or long-slit studies specifically targeted brightest cluster galaxies (BCGs) or galaxies in dense environments (e.g.,
\citealt{Loubseretal2008, Broughetal2011, Jimmyetal2013, Loubseretal2018, Krajnovicetal2018}).

A key result from these studies is the realization that the stellar kinematic properties of local ETGs depend strongly on the stellar mass $M_*$.  
At $M_* \la 10^{11} M_\odot$, around 90\% of the ETGs are fast rotators with a spin parameter above $\sim 0.2$, regular velocity features, aligned kinematic and photometric axes, and intrinsic axisymmetry (e.g., \citealt{Emsellemetal2007, Krajnovicetal2011, Weijmansetal2014, Fogartyetal2015, Broughetal2017, Fosteretal2017}). At $M_* \ga 10^{11.5} M_\odot$, however, the ETGs
become predominantly slow rotators with diverse kinematic features, misaligned kinematic and photometric axes, and intrinsic triaxiality (e.g., \citealt{Vealeetal2017b,Vealeetal2017a, Eneetal2018, Krajnovicetal2018}).

All the aforementioned IFS studies with the exception of \citet{Krajnovicetal2018} probed galaxy kinematic properties above $1''$ due to the limited spatial sampling scale of the instruments, e.g., $1.6''$, $2''$, and $2.7''$ for the fiber diameter of SAMI, MaNGA, and CALIFA, respectively, and $0.94''$ for the lenslet size of SAURON/ATLAS$^{\rm 3D}$. 
For a subsample of 18 ETGs in the SAURON survey,
\citet{McDermidetal2006} observed the central $8'' \times 10''$ region using the OASIS IFS with a spatial sampling of $0.27''$. 
These lower-mass ETGs ($M_* \sim 10^{10} - 10^{11.6}M_\odot$) are mainly fast rotators and many show emission lines. Their finely-resolved velocity maps revealed different types of kinematically distinct central components.

We designed the MASSIVE survey to study
massive ETGs with $M_* > 10^{11.5}M_\odot$ located within 108 Mpc in the northern sky through a combination of high angular resolution and wide-field IFS and photometric observations \citep{Maetal2014}.
We presented the wide-field kinematics measured from IFS observations taken over a $107'' \times 107''$ FOV in \cite{Vealeetal2017a, Vealeetal2017b, Vealeetal2018} and \cite{Eneetal2018}.
The latest MASSIVE paper \citep{Eneetal2019} presented the first results from the high angular resolution part of the survey  using the Gemini Multi Object Spectrograph (GMOS; \citealt{Hooketal2004}) on the Gemini North telescope.  
With a $5'' \times 7''$ field of view (FOV) and $0.2''$ spatial sampling, we obtained unprecedentedly detailed kinematic maps of the central $\sim 0.2$ kpc to 2 kpc regions of 20 MASSIVE galaxies.
We found a prevalence of positive $h_4$ and rising velocity dispersion profiles towards small radii indicative of central black holes and velocity dispersion anisotropy at the center of these massive ETGs.

This paper is the second of the high-resolution MASSIVE papers in which we focus on the velocity profiles and kinematic axes of the same 20 MASSIVE galaxies as in \citet{Eneetal2019}.
We use the kinemetry method of \cite{Krajnovicetal2006} to measure the misalignment between the kinematic axis and the photometric major axis and quantify substructures found in the velocity maps such as kinematic twists or kinematically distinct components.

The paper is structured as follows.
In Section~\ref{sec:observations_and_data} we describe the galaxy sample and IFS observations.
In Section~\ref{sec:kinemetry_analysis} we explain how we derive the main kinematic properties such as the average and local kinematic position angles and the misalignment angles.
Section~\ref{sec:central_rotation} presents results for the velocity amplitudes and kinematic axes in both the central regions and the main bodies of the sample galaxies and examines the misalignments of the central kinematic axis, the main-body kinematic axis, and the photometric major axis.
We analyze the local radially-dependent kinematic features in Section~\ref{sec:kinematic_radial_profiles} and discuss six galaxies with distinctive kinematic features in Section~\ref{sec:distinct_kinematic_features}.
 In Section~\ref{sec:discussion} we discuss 
 the assembly pathways for local massive ETGs in the broader context of numerical simulations. 
Section~\ref{sec:summary} summarizes our findings.


\section{Observations and Data}
\label{sec:observations_and_data}

In this paper we study the detailed velocity features of the central $\sim 2$ kpc of 20 galaxies in the MASSIVE survey \citep{Maetal2014}.
The list of 20 galaxies and their properties are given in Table~\ref{tab:properties}.
The galaxies are located between 54 Mpc and 102 Mpc distance (with a median distance of 70 Mpc) and all have stellar masses above $\sim 10^{11.7} M_\odot$.
The individual distances and $K$-band absolute magnitudes are listed in columns 2 and 3 of Table~\ref{tab:properties}.
Here we provide a brief description of the observations and data reduction procedures; an in-depth discussion is given in \cite{Eneetal2019}.

\subsection{High-angular resolution IFS observations}

We observe the central $5'' \times 7''$ region of each galaxy with the GMOS IFS on the 8.1 m Gemini North telescope.
Each galaxy is observed multiple times in order to meet a minimum signal-to-noise ratio (S/N).
The total exposure times range from 1 to 6 hrs, with most galaxies being observed for 3 hrs, on average.
Each science exposure provides one spectrum per lenslet for each of the 1000 lenslets of $0.2''$ spatial resolution.
An additional 500 lenslets observe an empty sky field with a $5'' \times 3.5''$ field of view (FOV) that is offset by $1'$ from the science field.
The spectra are in the wavelength range 7800 -- 9330 \AA~and have an average spectral resolution of 2.5 \AA~full width at half maximum (FWHM).

We follow the standard GMOS data reduction procedure using the Gemini package within the image reduction and analysis facility (IRAF) software to obtain wavelength-calibrated and sky-subtracted one-dimensional spectra for each spatial position on the galaxy.
We implement custom built routines to extract and combine the spectra from multiple exposures and spatially bin the data to S/N $\sim 120$ using the Voronoi binning routine of \cite{CappellariCopin2003}.
The binning process generates between 50 and 300 spatial bins, depending on the galaxy, with an average of $\sim 130$ bins per galaxy.

We use the penalized pixel-fitting (pPXF) routine of \cite{CappellariEmsellem2004} to measure the stellar LOSVD.
This method convolves the observed galaxy spectrum with a set of stellar templates to obtain the best-fitting LOSVD which is modelled as a Gauss-Hermite series of order $n=6$ (\citealt{Gerhard1993, vanderMarelFranx1993}):

\begin{equation}
\label{eq:GH}
f(v) = \frac{e^{-\frac{y^2}{2}}}{\sqrt{2 \pi \sigma^2}} \bigg[1 + \sum_{m=3}^n h_m H_m(y) \bigg],
\end{equation}

\noindent where $y=(v-V)/\sigma$, $V$ is the mean velocity, $\sigma$ is the velocity dispersion, and $H_m$ is the $m^{\text{th}}$ Hermite polynomial (using the definition in Appendix A of \cite{vanderMarelFranx1993}).
For more details on the optimal pPXF parameters for GMOS data and running the fitting procedure, see \cite{Eneetal2019}.

\subsection{Wide-field IFS observations}

The wide-field IFS data for 90 MASSIVE galaxies (which includes the 20 GMOS galaxies) are presented in \cite{Vealeetal2017a,Vealeetal2017b}.
The observations were taken with the Mitchell/VIRUS-P IFU at the 2.7-m Harlam J. Smith Telescope at McDonald Observatory, which has a FOV of $107'' \times 107''$ and a spatial sampling of $4.1''$.
The spectra cover the wavelength range 3650 -- 5850 \AA~with 5 \AA~FWHM average spectral resolution.
Full details of the observing strategy and data reduction procedure are given in \cite{Maetal2014} and \cite{Vealeetal2017a}.
While \cite{Vealeetal2017a} present 'folded' maps of the kinematic moments $V, \sigma, h_3-h_6$ (i.e in order to increase S/N the spectra are folded across the major photometric axis prior to binning), in \cite{Eneetal2018} we used the Mitchell IFS data to generate 'unfolded' maps of the LOSVD moments.
We then ran kinemetry on the unfolded velocity maps to measure the misalignment between the large scale ($\sim 1 R_e$) kinematics and photometry and presented radial profiles for several kinemetry coefficients.

\subsection{Photometric data}

We measured the surface brightness profiles and isophotal properties for a sample of 35 MASSIVE galaxies using observations taken with the Infrared Channel of the HST Wide Field Camera 3 (WFC3) in \cite{Goullaudetal2018}.
For 18 of the 20 galaxies studied here, we use the average photometric position angle, PA$_{\rm phot}$, determined in that work; the values and formal uncertainties are quoted in column 9 of Table~\ref{tab:properties}.
The galaxies NGC 2340 and NGC 4874 were not targeted by our program because they have archival HST observations.
For these two galaxies, we use the PA$_{\rm phot}$ values from the Two Micron All Sky Survey (2MASS; \citealt{Skrutskieetal2006}) catalogue for NGC~2340 and from the NASA-Sloan Atlas (NSA; \citealt{Yorketal2000,Aiharaetal2011}) for NGC~4874.
Neither catalogue provides uncertainties for the photometric PA, so we assume a fiducial error of $5\degr$.

Most galaxies in our sample show fairly regular photometric profiles where the isophotal position angle changes by less than $15\degr$ across the radial extent of the WFC3 data (from $\sim 0.2$ kpc to $\sim 20$ kpc).
A handful of galaxies have more complex photometric profiles, usually showing more pronounced isophotal twists of greater than $20\degr$.
Among these galaxies, NGC~1129 has the most interesting photometric profile.
\cite{Goullaudetal2018} report the luminosity-weighted average photometric PA (computed using all available isophotes) to be $61.7\degr$, but the detailed radial profile shows two distinct regions of constant PA: an inner component within $\sim 10''$ with PA $\sim 0\degr$ and an outer region with PA $\sim 90\degr$.
In order to provide a fair comparison with the GMOS kinematics, we report PA$_{\rm phot}$ using only isophotes corresponding to the inner component within $\sim 10''$.

We also report the half-light radius $R_e$ measured from deep K-band photometric data taken with WIRCam on the Canada-France-Hawaii Telescope (CFHT) as part of the MASSIVE survey (M. E. Quenneville et al., in preparation). 
The effective radius is measured using the photometry package ARCHANGEL \citep{Schombert2007} which fits elliptical isophotes to the stacked image of each galaxy. 
The aperture luminosity for each isophote (as a function of radius) is used to construct a curve of growth.
The total luminosity and half-light radius are then measured from the curve of growth. 
The values of $R_e$ for our sample of 20 galaxies are given in column 4 of Table~\ref{tab:properties}. 
They range from $\sim 5$ to $\sim 20$ kpc, with the average $R_e$ being $\sim 9$ kpc.


\section{Kinemetry Analysis}
\label{sec:kinemetry_analysis}

To analyze the velocity maps of our sample galaxies we use the \texttt{kinemetry}\footnote{http://davor.krajnovic.org/idl/} method \citep{Krajnovicetal2006}.
In the following sections we describe how we apply this method to measure global (Section~\ref{sub:global_kinematic_position_angle}) and local (Section~\ref{sub:spatially_resolved_velocity_profiles}) kinematic parameters.

\begin{deluxetable*}{lrccccccccccc}
\tablecaption{Galaxy properties and kinemetry derived quantities for the 20 MASSIVE galaxies. \label{tab:properties}}
\tablehead{
\colhead{Galaxy} & \colhead{$D$} & \colhead{$M_K$} & \colhead{$R_e$} & \colhead{$\lambda_{\rm 1~kpc}$} & \colhead{$\lambda_{e}$} & \colhead{PA$_{\rm kin}^{\rm GMOS}$} & \colhead{PA$_{\rm kin}^{\rm Mitchell}$} & \colhead{PA$_{\rm phot}$} & \colhead{$\Psi_{\rm GMOS}$} & \colhead{$k_{1, {\rm GMOS}}^{\rm max}$} & \colhead{$k_{1, {\rm Mitchell}}^{\rm max}$} & \colhead{Env} \\
\colhead{} & \colhead{[Mpc]} & \colhead{[mag]} & \colhead{[kpc]} & \colhead{} & \colhead{} & \colhead{[deg]} & \colhead{[deg]} & \colhead{[deg]} & \colhead{[deg]} & \colhead{[\kms]} & \colhead{[\kms]} & \colhead{}
}
\colnumbers
\startdata
NGC0057 &  76.3 & $-25.75$ & 6.31 & 0.025 & 0.028 & 100 $\pm$ 22 & -- &  40.2 $\pm$ 0.5 & 59.3 $\pm$ 22.0 &   9 & 10 & I\\
NGC0315 &  70.3 & $-26.30$ & 9.20 & 0.027 & 0.063 & 218 $\pm$ 13 & 222 $\pm$ 7 &  44.3 $\pm$ 0.2 &  6.3 $\pm$ 13.3 &  23 & 44 & B \\
NGC0410 &  71.3 & $-25.90$ & 7.57 & 0.052 & 0.048 & 211 $\pm$  9 & 161 $\pm$ 19 &  35.8 $\pm$ 0.9 &  4.8 $\pm$  9.3 &  29 & 19 & B \\
NGC0545 &  74.0 & $-25.83$ & 9.71 & 0.034 & 0.081 & 287 $\pm$ 17 & -- &  57.2 $\pm$ 0.7 & 49.8 $\pm$ 17.3 &  13 & 11 & B \\
NGC0547 &  71.3 & $-25.90$ & 10.55 & 0.024 & 0.081 & 254 $\pm$ 31 & -- &  98.8 $\pm$ 1.4 & 24.8 $\pm$ 31.0 &   9 & 30 & S \\
NGC0741 &  73.9 & $-26.06$ & 9.74 & 0.037 & 0.050 & 202 $\pm$ 16 & -- &  88.0 $\pm$ 1.1 & 66.5 $\pm$ 16.3 &  15 & 12 & B\\
NGC0777 &  72.2 & $-25.94$ & 5.89 & 0.027 & 0.060 & 311 $\pm$ 22 & 8 $\pm$ 10 & 148.6 $\pm$ 0.8 & 18.1 $\pm$ 21.5 &  12 & 41 & B\\
NGC0890 &  55.6 & $-25.50$ & 6.62 & 0.027 & 0.014 & 159 $\pm$ 42 & 101 $\pm$ 9 &  53.7 $\pm$ 0.3 & 74.6 $\pm$ 42.3 &   9 & 46 & I \\
NGC1016 &  95.2 & $-26.33$ & 9.47 & 0.015 & 0.040 & -- & 262 $\pm$ 20 & 42.8 $\pm$ 1.0 & -- & 7 & 30 & B \\
NGC1060 &  67.4 & $-26.00$ & 6.38 & 0.034 & 0.048 & 351 $\pm$ 10 & 342 $\pm$ 14 &  74.8 $\pm$ 0.4 & 83.8 $\pm$ 10.0 &  25 & 15 & B\\
NGC1129 &  73.9 & $-26.14$ & 16.13 & 0.350 & 0.124 & 185 $\pm$ 1 & 179 $\pm$ 6 &  7.7 $^\dagger$ $\pm$ 0.5 & 3.2 $\pm$  1.1 & 148 & 66 & B \\
NGC1453 &  56.4 & $-25.67$ & 6.00 & 0.199 & 0.204 & 25 $\pm$ 3 & 35 $\pm$ 3 &  30.1 $\pm$ 0.2 &  5.1 $\pm$  3.3 &  99 & 92 & B \\
NGC1573 &  65.0 & $-25.55$ & 5.43 & 0.026 & 0.056 & 181 $\pm$ 53 & 190 $\pm$ 19 &  31.7 $\pm$ 0.1 & 31.2 $\pm$ 53.3 &   7 & 26 & B \\
NGC1600 &  63.8 & $-25.99$ & 9.14 & 0.045 & 0.035 &  18 $\pm$  4 & -- &   8.8 $\pm$ 0.1 &  8.7 $\pm$  4.3 &  18 & 22 & B \\
NGC1700 &  54.4 & $-25.60$ & 4.45 & 0.119 & 0.198 &  87 $\pm$ 2 & 268 $\pm$ 2 &  90.6 $\pm$ 0.3 &  3.6 $\pm$  1.8 &  51 & 115 & B \\
NGC2258 &  59.0 & $-25.66$ & 5.76 & 0.034 & 0.071 & 74 $\pm$  9 & 71 $\pm$ 17 & 150.8 $\pm$ 1.2 & 77.3 $\pm$  8.6 &  19 & 31 & B \\
NGC2274 &  73.8 & $-25.69$ & 6.57 & 0.042 & 0.073 & 231 $\pm$ 7 & 288 $\pm$ 26 & 165.0 $\pm$ 0.2 & 66.0 $\pm$  7.0 &  21 & 26 & B \\
NGC2340 &  89.4 & $-25.90$ & 14.27 & 0.042 & 0.032 & 53 $\pm$  6 & -- &  80.0 $\pm$ 5.0 & 27.0 $\pm$  7.6 &  18 & 12 & S \\
NGC2693 &  74.4 & $-25.76$ & 5.63 & 0.337 & 0.294 & 172 $\pm$ 1 & 169 $\pm$ 2 & 161.3 $\pm$ 1.3 & 10.7 $\pm$  1.8 & 157 & 144 & I \\
NGC4874 & 102.0 & $-26.18$ & 19.20 & 0.018 & 0.072 & -- & 335 $\pm$ 6 &  40.6 $\pm$ 5.0 & -- & 9 & 40 & S \\
\enddata
\tablecomments{
    (1) Galaxy name.
    (2) Distance from Paper I \citep{Maetal2014}.
    (3) Absolute $K$-band magnitude from Paper I \citep{Maetal2014}.
    (4) Effective radius from CFHT deep K-band photometry (M. E. Quenneville et al., in preparation).
    (5) Spin parameter within 1 kpc measured from GMOS IFS data reported in  Paper XIII \citep{Eneetal2019}.
    (6) Spin parameter within one effective radius measured from Mitchell IFS data reported in Paper X \citep{Eneetal2018}.
    (7) Kinematic position angle (measured E of N to the receding part) within the FOV of the GMOS IFS. See Section~\ref{sub:global_kinematic_position_angle} for details.
    (8) Kinematic position angle (measured E of N to the receding part) within the FOV of the Mitchell IFS reported in Paper X \citep{Eneetal2018}. See Section~\ref{sub:global_kinematic_position_angle} for details.
    (9) Luminosity-weighted average photometric position angle from Paper IX \citep{Goullaudetal2018}. 
    $^\dagger$ The photometric PA for NGC~1129 shows a prominent twist beyond $\sim 10''$; the quoted value here is measured within $10''$. See Fig.~7 and Sections 2.3 and 6.3 for details. 
    The photometric PAs for NGC~2340 and NGC~4874 are taken from 2MASS and NSA, respectively.
    (10) Misalignment angle between GMOS kinematic axis and the photometric major axis. See Section~\ref{sub:global_kinematic_position_angle} for details.
    (11) Maximum value of the velocity coefficient $k_1$ measured within the GMOS FOV ($R \sim 1$ kpc).
    (12) Maximum value of the velocity coefficient $k_1$ measured within the Mitchell FOV ($R \sim 1 R_e$).
    (13) Galaxy environmental types according to the 2MASS group catalog from Paper I \citep{Maetal2014}: B for brightest group or cluster galaxy; S for satellites; I for isolated galaxies.
    }
\end{deluxetable*}

\subsection{Global kinematic position angle}
\label{sub:global_kinematic_position_angle}

To identify any coherent kinematic structure in the GMOS velocity maps, we measure the global kinematic position angle, PA$_{\rm kin}^{\rm GMOS}$, using the \texttt{fit\_kinematic\_pa}\footnote{http://www-astro.physics.ox.ac.uk/{\raise.17ex\hbox{$\scriptstyle\sim$}}mxc/software/} routine described in Appendix C of \cite{Krajnovicetal2006}.
Briefly, the routine generates a bi-antisymmetric model velocity map for each possible value of kinematic position angle and compares it to the observed velocity map.
The reported PA$_{\rm kin}$ corresponds to the best-fitting model that minimizes the $\chi^2$ between the observed and model velocity maps.
The routine also assigns error estimates to the best-fit PA$_{\rm kin}$ as the range of angles for which $\Delta \chi^2 < 1$, which corresponds to the $1\sigma$ confidence level for one parameter.
The reported error bars anti-correlate with the amount of organized rotation: in cases with strong rotation, the error bars are very tight, while for cases with little or no organized rotation, the error bars approach $90\degr$.

The global kinematic position angle PA$_{\rm kin}^{\rm GMOS}$ thus gives the average direction of rotation in the central few kpc of each galaxy. It is measured counterclockwise from north to the receding part of the galaxy within the GMOS FOV.
The determination of PA$_{\rm kin}^{\rm GMOS}$ enables us to measure the relative alignment angle between the GMOS kinematic axis and the photometric axis, following the convention of \cite{Franxetal1991}:
\begin{equation}
\label{eqn:psi}
    \sin \Psi_{\rm GMOS} = |\sin ({\rm PA_{kin}^{GMOS}} - {\rm PA_{phot}})| \,.
\end{equation}
The error bars on $\Psi_{\rm GMOS}$ are computed as the quadrature sum of the errors on ${\rm PA_{kin}^{GMOS}}$ and ${\rm PA_{phot}}$.
Our measurements of ${\rm PA_{kin}^{GMOS}}$ and $\Psi_{\rm GMOS}$ are tabulated in Table~\ref{tab:properties} and discussed in Section~\ref{sub:misalignment_between_kinematics_and_photometry}.

\subsection{Spatially-resolved velocity profiles}
\label{sub:spatially_resolved_velocity_profiles}

\begin{figure*}
    \gridline{\fig{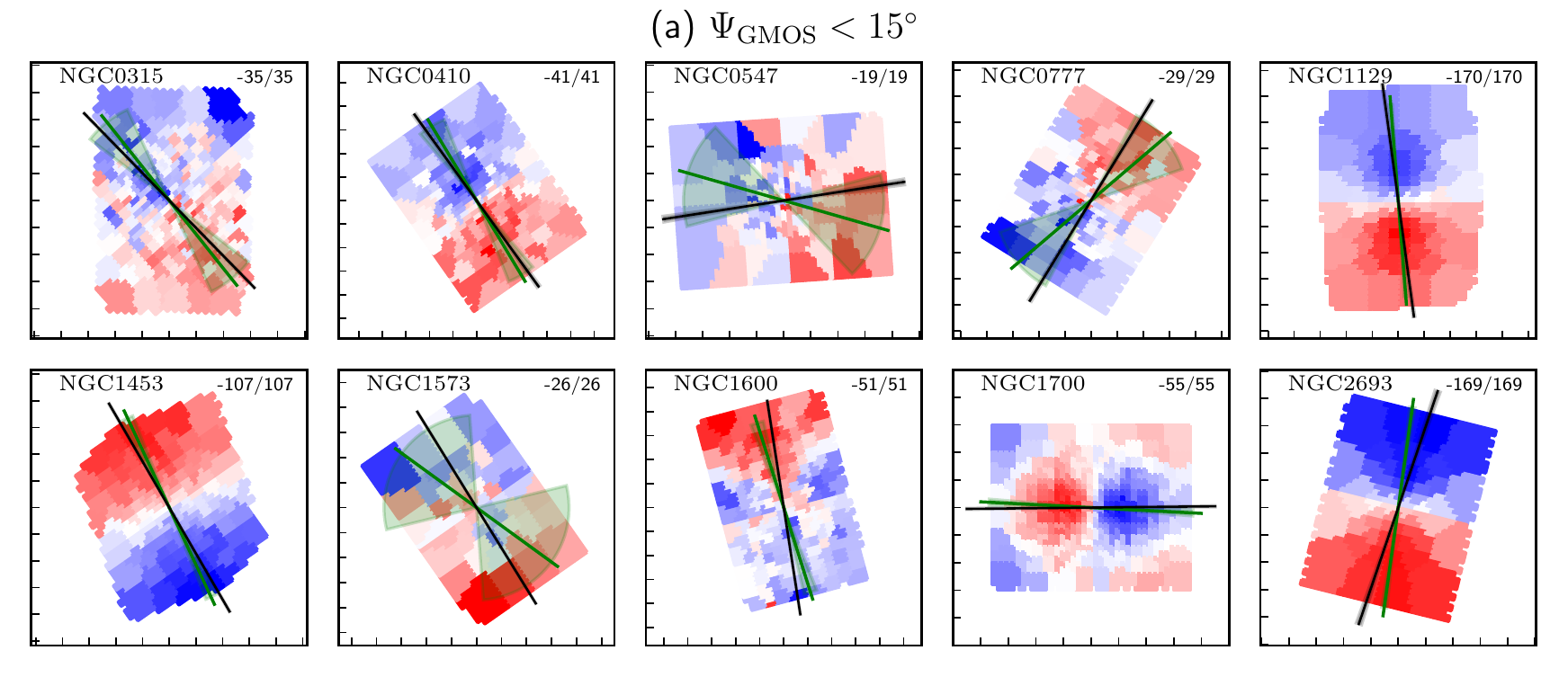}{\textwidth}{}
    }
    \vspace{-1cm}
    \gridline{\fig{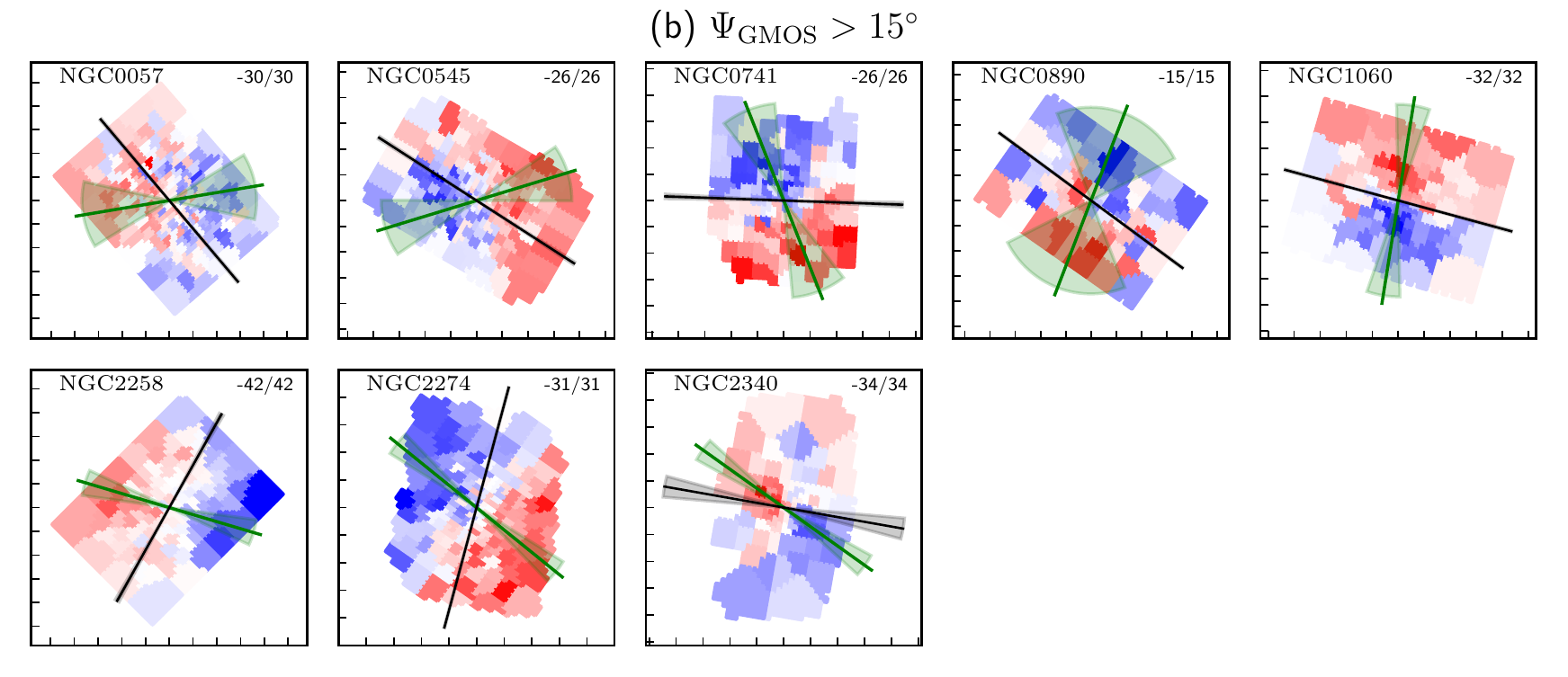}{\textwidth}{}
    }
    \vspace{-1cm}
    \gridline{\fig{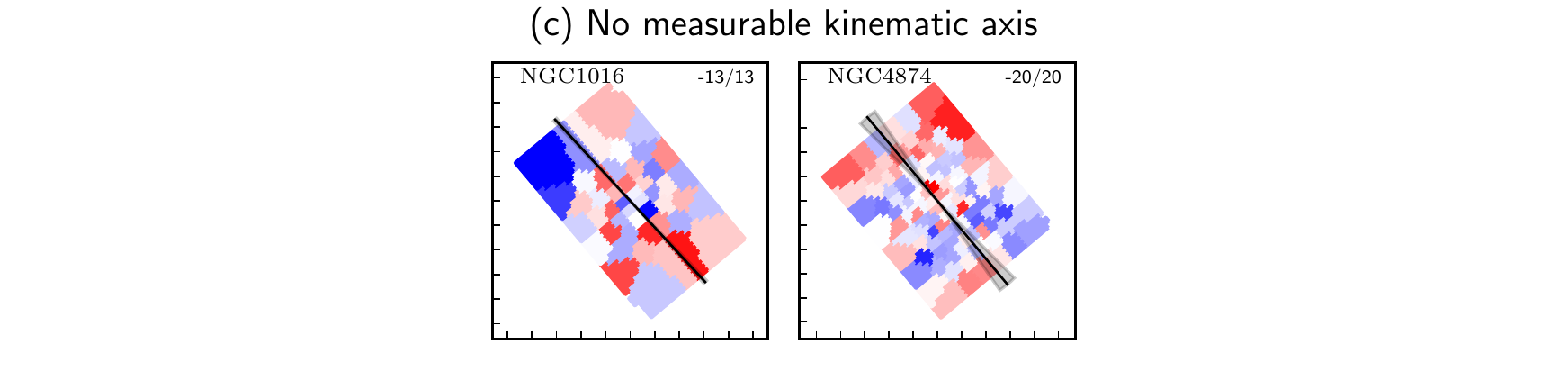}{\textwidth}{}
    }
    \caption{GMOS velocity maps of the central $5''\times 7''$ region of 20 MASSIVE galaxies studied in this paper, grouped by the misalignment angle \psiG\ between the central kinematic axis (green line) and photometric axis (black line):
   (a) \psiG\ $< 15^\circ$, (b) \psiG\ $> 15^\circ$, and (c) no measurable kinematic axis.
    The green and black wedges represent the $1 \sigma$ uncertainties in the kinematic PA, \PAkinG, and the photometric PA, respectively.
    Red and blue colors denote receding and approaching velocities, respectively.
    The color scale in each map corresponds to the velocity range (in \kms) given in the upper right corner of each subpanel.
    Tickmarks are spaced by $1''$ in each subpanel.
	The photometric PA for NGC~1129 shows a prominent twist beyond $\sim 10''$; the quoted value here is measured within $10''$. See Fig.~7 and Sections 2.3 and 6.3 for details. 
             }
    \label{fig:test}
\end{figure*}

We measure the local kinematic features in the GMOS velocity maps using the \texttt{kinemetry} method of \cite{Krajnovicetal2006}.
Kinemetry is an extension of isophotal analysis which models the maps of LOSVD moments as simple functional forms along ellipses: a constant for symmetric (even) moments and a cosine term for antisymmetric (odd) moments.
In particular, kinemetry uses Fourier decomposition to model the velocity profile along an ellipse as the sum of $N+1$ harmonic terms:
\begin{equation}
\label{eqn:kinemetry}
V(a,\psi) = A_0(a) + \sum_{n=1}^N k_n(a) \cos[n(\psi - \phi_n(a))],
\end{equation}
where $a$ is the length of the semi-major axis of the ellipse, $\psi$ is the eccentric anomaly, $A_0$ is related to the systemic velocity of the galaxy, and $k_n$ and $\phi_n$ are the amplitude and phase coefficients, respectively. 	
The leading term $k_1$ represents the amplitude of the rotational motion. The coefficient
$k_5$ represents higher-order deviations from the simple cosine law assumption, where a high $k_5$ value indicates the presence of multiple kinematic components.

The main outputs of the kinemetry code are the kinematic coefficients $k_n$, and two geometry coefficients that specify the local position angle $\Gamma$ and the flattening $q_{\rm kin}$ of the best-fitting ellipses along which velocity extraction was performed ($q_{\rm kin}=1$ corresponds to velocity extraction along circles).
The code determines these parameters in two steps. 
In the first step, a kinematic profile is extracted at each radius $a$ for each value of ($\Gamma$, $q_{\rm kin}$) chosen from a finely-sampled grid. 
The best-fitting  $\Gamma$ and $q_{\rm kin}$ are the ones found to minimize $\chi^2 = k_1^2 + k_2^2 + k_3^2$.
Then, in the second step, the kinematic coefficients $k_n$ are computed through Fourier decomposition along the ellipse given by the best-fitting $\Gamma$ and $q_{\rm kin}$ of the previous step.

For our sample, we first let kinemetry perform velocity extraction along best-fitting ellipses.
This is well suited for the handful of galaxies with high velocity gradients within the GMOS FOV: NGC 1129, NGC 1453, NGC 1700, and NGC 2693. 
Applying kinemetry to galaxies that rotate much slowly (the majority of our sample), however, is more complicated since the low velocity gradients introduce significant degeneracies between the position angle and flattening parameters. 
In order to reduce the degeneracy in such cases, we opt to rerun kinemetry along best-fitting circles (i.e., setting $q_{\rm kin}=1$). 


\begin{figure*}
    \plottwo{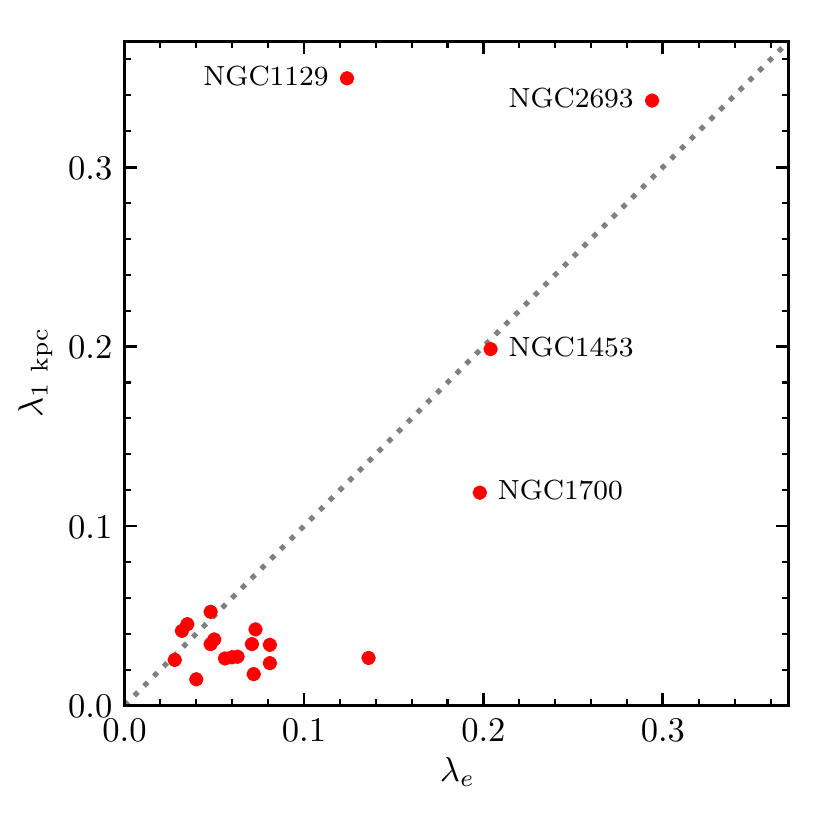}{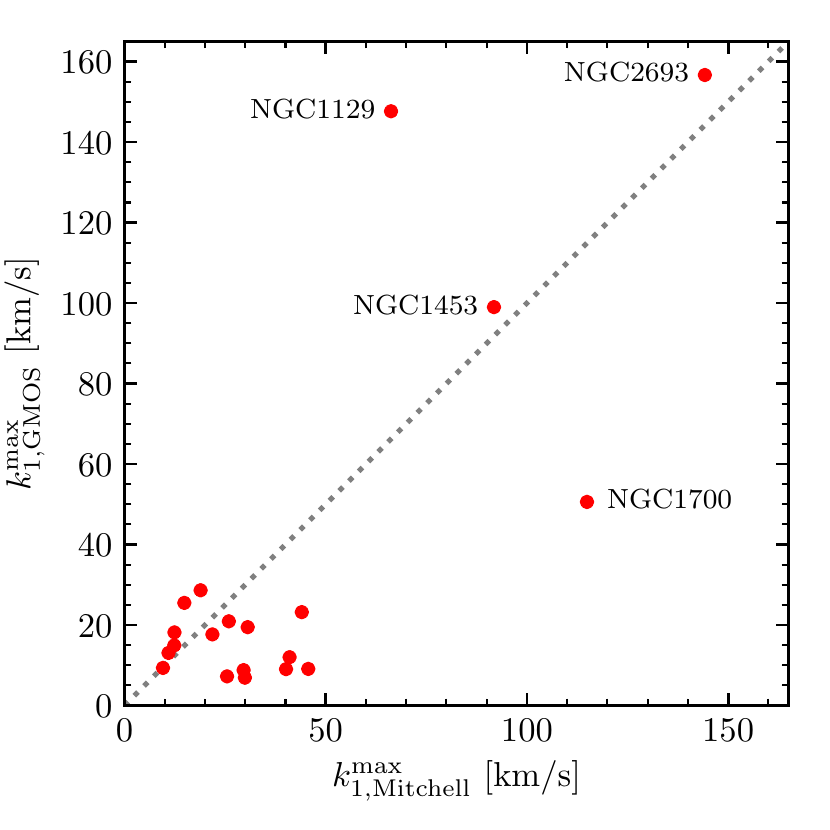}
    \caption{Amount of rotation in the central few kpc versus within an effective radius, as measured by the spin parameter $\lambda$ (left panel) and by the peak value of the velocity coefficient $k_{1}^{\rm max}$ (right panel) for the 20 MASSIVE galaxie studied in this paper.
    Four galaxies --  NGC~1129, NGC~1453, NGC~1700, and NGC~2693 -- have both high central and high global rotations.  They are clearly separated from the rest of the sample, which
     clusters in the lower left corner with $\lambda$ below 0.1 and $k_{1}^{\rm max}$ below 50 \kms.
    In both panels the gray dotted line indicates the one-to-one relation.
    }
    \label{fig:maxk1_gmos_vp}
\end{figure*}

\section{Central and Main-body Rotation}
\label{sec:central_rotation}

The finely-resolved GMOS velocity map of the central few kpc for each of the 20 MASSIVE galaxies is shown in Figure~\ref{fig:test}.  Below we discuss the amplitudes and axes of the detected rotations and analyze how the kinematic axis in the central region of each galaxy is oriented relative to its photometric major axis and its main-body rotation measured within an effective radius ($\sim 10$ kpc).

\subsection{Amplitude of rotation}
\label{sub:amplitude_of_rotation}

As a measure of the importance of rotation compared to velocity dispersion in each galaxy, we determine the spin parameter $\lambda$ within a circular aperture of radius $R$, defined as $\lambda(<R) \equiv \langle R|V| \rangle / \langle R\sqrt{V^2 + \sigma^2} \rangle $.
The brackets denote luminosity-weighted average quantities.
The spin parameters measured within 1 kpc from our GMOS data, $\lambda_{\rm 1~kpc}$, are listed in column 5 of Table~\ref{tab:properties} and plotted in the left panel of Figure~\ref{fig:maxk1_gmos_vp}. 
These values are compared to the main-body spins measured within one effective radius, $\lambda_e$, from our Mitchell IFS data (\citealt{Eneetal2018}; column 6 of Table~\ref{tab:properties}).

All but four galaxies have low central spins as well as low main-body spins.  
Our earlier analysis of the main-body rotation in 370 galaxies in the MASSIVE and ATLAS$^{\rm 3D}$ surveys  found a strong dependence of $\lambda_e$ on stellar mass, where the mean $\lambda_e$ declines from $\sim 0.4$ at $M_*\sim 10^{10} M_\odot$ to below 0.1 at $M_*\sim 10^{12} M_\odot$ \citep{Vealeetal2017b}. 
Figure~\ref{fig:maxk1_gmos_vp} shows that the low spin continues to the core in the majority of ETGs in the high mass regime.

As another measure of rotation, the right panel of Figure~\ref{fig:maxk1_gmos_vp} shows the maximum value of the velocity coefficient $k_1$ defined in Eq.~(\ref{eqn:kinemetry}). 
The individual values of $k_{\rm 1, GMOS}^{\rm max}$ and $k_{\rm 1, Mitchell}^{\rm max}$ are listed in columns 11 and 12 of Table~\ref{tab:properties}, respectively.
The results are very similar to those in the left panel of Figure~\ref{fig:maxk1_gmos_vp}, where the same 16 galaxies with spins below $\sim 0.1$ also have $k_1^{\rm max} \lesssim 50$ \kms\ over the entire radial range of $\sim 0.2$ kpc to 20 kpc covered by our IFS data.

The four highlighted galaxies in Figure~\ref{fig:maxk1_gmos_vp} have significantly higher spin and peak velocity than the rest of the sample.  However, only two of them -- NGC~1453 and NGC~2693 -- lie along the diagonal in Figure~\ref{fig:maxk1_gmos_vp} and are regular fast rotators in which
the central part co-rotates with the main body of the galaxy.
The other two galaxies -- NGC~1129 and NGC~1700 -- are unusual and have different central and main-body rotations.
As we will discuss further in Section~\ref{sec:distinct_kinematic_features},
the central region of NGC~1700 rotates in exactly the opposite direction as the main body, while the photometric PA in the central part of NGC~1129 twists by a striking $90\degr$ relative to the main body. 

\subsection{Axis of rotation: kinematic versus photometric PA}
\label{sub:misalignment_between_kinematics_and_photometry}

\begin{figure}
    \includegraphics[width=\columnwidth]{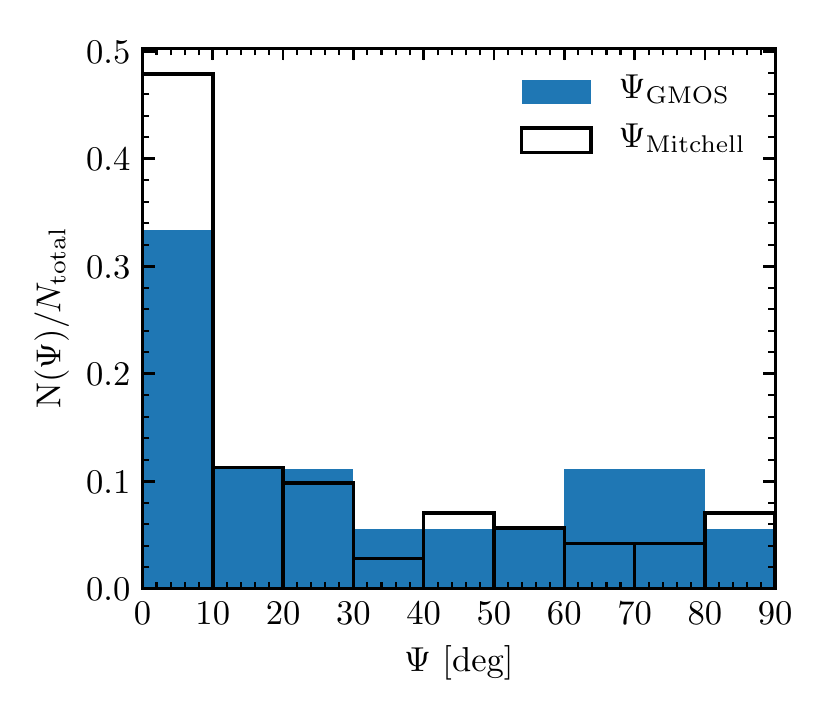}
    \caption{
    Histogram of the misalignment angle between the kinematic and photometric axes measured from GMOS IFS data for 18 galaxies with identifiable rotations in the central few kpc.  Within this sample,
    7 galaxies show aligned rotation with \psiG\ $\lesssim 15\degr$ and 11 galaxies show misaligned rotation that varies from mild misalignment to maximal misalignment of $90\degr$.
    The distribution of the misalignment angle measured from the large-scale Mitchell IFS data for 71 MASSIVE galaxies published in \citet{Eneetal2018}  is overplotted in black.  The two distributions show similarly flat and long tails in $\Psi$.}
    \label{fig:gmos_psi_hist}
\end{figure}

For 18 of the 20 galaxies in the sample,\footnote{For two galaxies, NGC 1016 and NGC 4874,
the algorithm could not find a well-defined kinematic axis, i.e., the $1\sigma$ errors on \PAkinG\ are $\sim 90\degr$.} we are able to identify a kinematic axis for the stellar rotation within the GMOS FOV using the algorithm described in Section~\ref{sub:global_kinematic_position_angle}.
The detected kinematic axes are represented by green lines in the GMOS velocity maps in Figure~\ref{fig:test}.  The photometric major axis for each galaxy is overplotted for comparison (black line).
Our measured values for the central kinematic PA ${\rm PA}_{\rm kin}^{\rm GMOS}$, photometric PA and the misalignment angle $\Psi_{\rm GMOS}$ (defined in Eq.~\ref{eqn:psi}) are given in columns 7, 9 and 10 in Table~\ref{tab:properties}, respectively.  

For NGC~1016 and NGC~4874, we do not detect any rotation in the GMOS maps shown in the bottom panel of Figure.~\ref{fig:test}.

The distribution of the misalignment angle $\Psi_{\rm GMOS}$ is plotted in Figure~\ref{fig:gmos_psi_hist}.
It peaks at small values that corresponds to the standard scenario in which the galaxy rotates around its minor axis, i.e., the kinematic axis is aligned with the photometric major axis.  
The distribution has a long and relatively flat tail extending to the maximum misalignment of $90\degr$, which corresponds to rotation around the major axis sometimes referred to as ``minor-axis rotation" or ``prolate-like rotation".
We classify as aligned rotators the 10 galaxies (56\% of the sample) that either have \psiG\ $< 15\degr$, or have \psiG $> 15\degr$ but \PAkinG\ and PA$_{\rm phot}$ agreeing within the (large) errors on the kinematic axis.
We classify the other 8 galaxies as misaligned rotators.

We note that even though NGC~1573 is classified as ``aligned", it is a borderline case.  The GMOS kinematic map shows interesting velocity structures in the inner $\sim 2''$ (Fig.~1) and significant local kinematic twists (Fig.~6 below), resulting in large errors in the luminosity-averaged ${\rm PA}_{\rm kin}^{\rm GMOS}$ within the GMOS FOV. Further discussion is given in Sec.~5.2.

We have previously used a similar procedure to determine the 
misalignment angle between the main-body kinematic axis and the photometric major axis from the wide-field Mitchell data for 90 MASSIVE galaxies \citep{Eneetal2018}.  
The distribution of this main-body misalignment angle,
\psiM, is plotted in Figure~\ref{fig:gmos_psi_hist} for comparison.
We note that while the Mitchell sample size is much larger than the GMOS sample, the shapes of the \psiM\ and \psiG\ distributions are qualitatively very similar.
The larger peak at small misalignment angle in the Mitchell data is primarily driven by the presence of a larger fraction of fast rotators: 22 of the 71 galaxies with measurable Mitchell kinematic axis are fast rotators (defined to have $\lambda_e > 0.2$) in \cite{Eneetal2018}, versus 3 fast rotators in the current sample of 18 galaxies with measurable GMOS kinematic axis.

Neither of our misalignment angle distributions shows a gap in the intermediate range of $\sim 25\degr - 55\degr$ as seen in the MUSE sample of 25 massive ETGs in dense environments \citep{Krajnovicetal2018}.  Furthermore, more than 1/3 of their galaxies show prolate-like rotations ($\Psi > 75 \degr$), while prolate-like rotators constitute only 20\% of our GMOS sample and 10\% of our Mitchell sample.  
These differences could be due to small-number statistics as well as differing galaxy environments for the two samples.  
The MUSE sample specifically targets ETGs in dense clusters and superclusters, whereas the MASSIVE sample is selected purely by $M_*$ and spans a wide range of galaxy environments, from brightest cluster galaxies to nearly isolated massive ETGs \citep{Vealeetal2017b,Vealeetal2017a}.
For the subsample of 20 galaxies studied here, the environmental types according to the 2MASS group catalog are given in column 13 of Table~1: 14 are brightest group or cluster galaxies, 3 are satellites in their respective groups, and 3 do not have neighbors.
It would be very interesting to further assess the role of environments on kinematic misalignments with a larger sample of massive ETGs.

In \cite{Eneetal2018} we found a strong correlation between the main-body kinematic misalignment and spin parameter in MASSIVE galaxies: 91\% of the fast rotators are aligned, whereas only 43\% of the slow rotators are aligned with \psiM\ below $15\degr$.
Despite the smaller sample here,
we find a very similar trend for the central kinematics: all four galaxies with high central spins (Fig.~2) are very well aligned, while only 43\% of the low-spin galaxies are aligned (Fig.~\ref{fig:test}(a)).  
Previous kinematic studies of lower mass ETGs were focused on main-body misalignment on scales of an effective radius and found similar trends: 83\% of the 62 fast rotators and 38\% of the 16 slow rotators are aligned in the SAMI survey \citep{Fogartyetal2015}, while
96\% of the 224 fast rotators and 56\% of the 36 slow rotators are aligned in the ATLAS$^{\rm 3d}$ survey \citep{Krajnovicetal2011}.  Very few galaxies in these surveys, however, are in the high $M_*$ range probed by the MASSIVE survey.

In \cite{Eneetal2018} we used the observed main-body misalignment and ellipticity distributions to infer the intrinsic shape distribution of the MASSIVE slow rotators, which was found to be mildly triaxial with mean axis ratios of $b/a = 0.88$ and $c/a = 0.65$.
A larger sample than the current 20 GMOS galaxies would be needed to perform a similar statistical analysis to infer the intrinsic shape distribution in the central kpc region of the galaxies.

\subsection{Axis of rotation: central versus main-body kinematic PA}
\label{sec:large_scale_kinematics}

\begin{figure}
    \includegraphics[width=\columnwidth]{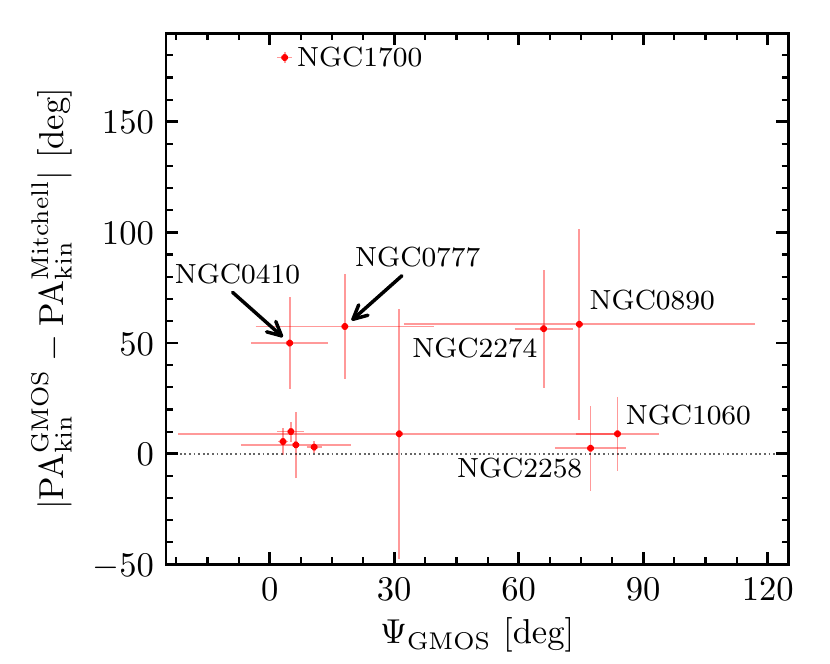}
    \caption{
    Alignment between the (average) kinematic axis in the central kpc (\PAkinG) and in the main body (\PAkinM) of 12 MASSIVE galaxies, plotted against the central kinematic misalignment angle (i.e., difference between \PAkinG\ and the photometric PA).
    Overall, 5 out of 12 galaxies show a significant difference ($\gtrsim 20\degr$) between the central and main-body rotation axes.
    The remaining 7 galaxies are all consistent with having the same average kinematic axis through the radial range probed by the two combined data sets.
    }
    \label{fig:pakin_scatter}
\end{figure}

In the previous two subsections we studied how the kinematic axes are aligned with the photometric axes.  Here we examine the alignment between the central and main-body kinematic axes
for the 12 galaxies that have sufficient rotations for the kinematic PAs to be determined in both GMOS and Mitchell observations (see Table~\ref{tab:properties}).

Figure~\ref{fig:pakin_scatter} shows the difference in the two kinematic PAs as a function of the central misalignment angle $\Psi_{\rm GMOS}$.
Overall, we find the central and main-body kinematic axes to be very well aligned in 7 of the 12 galaxies ($|{\rm PA}_{\rm kin}^{\rm GMOS} - {\rm PA}_{\rm kin}^{\rm Mitchell}| \la 10\degr$).  It is interesting to note, however, that only 5 of these galaxies also have aligned kinematic and photometric axes with small $\Psi$ (i.e., lower-left corner of Fig,~\ref{fig:pakin_scatter}).   
These 5 objects (NGC~315, NGC~1129, NGC~1453, NGC~1573, and NGC~2693) are the most aligned galaxies in our sample and show regular (albeit frequently slow) rotation about the minor photometric axis, as is seen for a large fraction of lower-mass early-type galaxies and disk galaxies.  
The other two galaxies, NGC~1060 and NGC~2258, have aligned central and main-body kinematic axes, but both are misaligned from the photometric major axis, and the misalignment is near orthogonal (see Table~1 and Fig.~1). 
We will discuss these ``minor-axis rotation" galaxies further in Sec.~\ref{sec:minor-axis}.
 
For the 5 of the 12 galaxies that have noticeably misaligned central and main-body kinematic axes in Figure~\ref{fig:pakin_scatter}  ($|{\rm PA}_{\rm kin}^{\rm GMOS} - {\rm PA}_{\rm kin}^{\rm Mitchell}| \ga 30\degr$), it is also interesting to ask whether either kinematic axis is aligned with the galaxy's photometric axis. We again find diverse behaviors even within this small sample. 
For two galaxies (NGC~410 and NGC~777), the central kinematic axis is well aligned with the photometric axis, but the main-body kinematic axis is not. Both galaxies have very regular isophotes with nearly constant photometric PA out to $\sim 100''$ in our HST WFC3 images (Figs.~8 and 16 of \citealt{Goullaudetal2018}). It is therefore the main-body rotation that show an  intermediate-angle ($\sim 50^\circ - 60^\circ$) misalignment. 
For two other galaxies (NGC~890 and NGC~2274), the three axes are all pointing in different directions.
This implies at least some level of intrinsic triaxiality since it is unlikely to have observed kinematic misalignment between all three axes that is due solely to projection effects.

The last galaxy, NGC~1700, is a special case in which the inner $\sim 1$ kpc is a kinematically distinct component that is counter-rotating with respect to the main body of the galaxy \citep{Eneetal2019}. We will discuss NGC~1700 further below.

\section{Local Kinematic Profiles}
\label{sec:kinematic_radial_profiles}

\begin{figure}
    \includegraphics[width=\columnwidth]{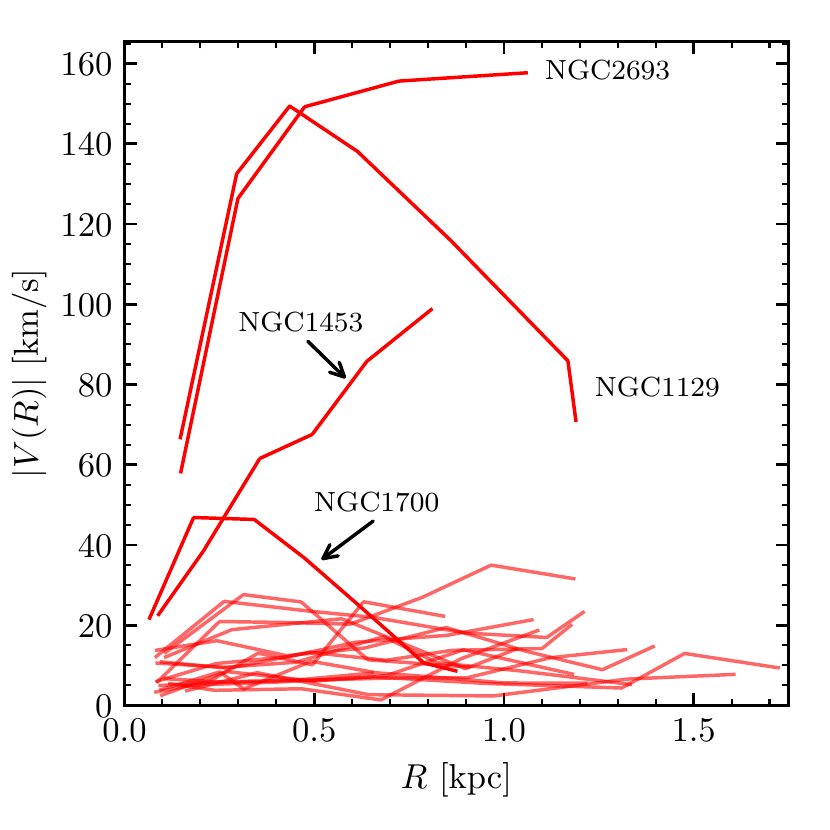}\\
    \includegraphics[width=\columnwidth]{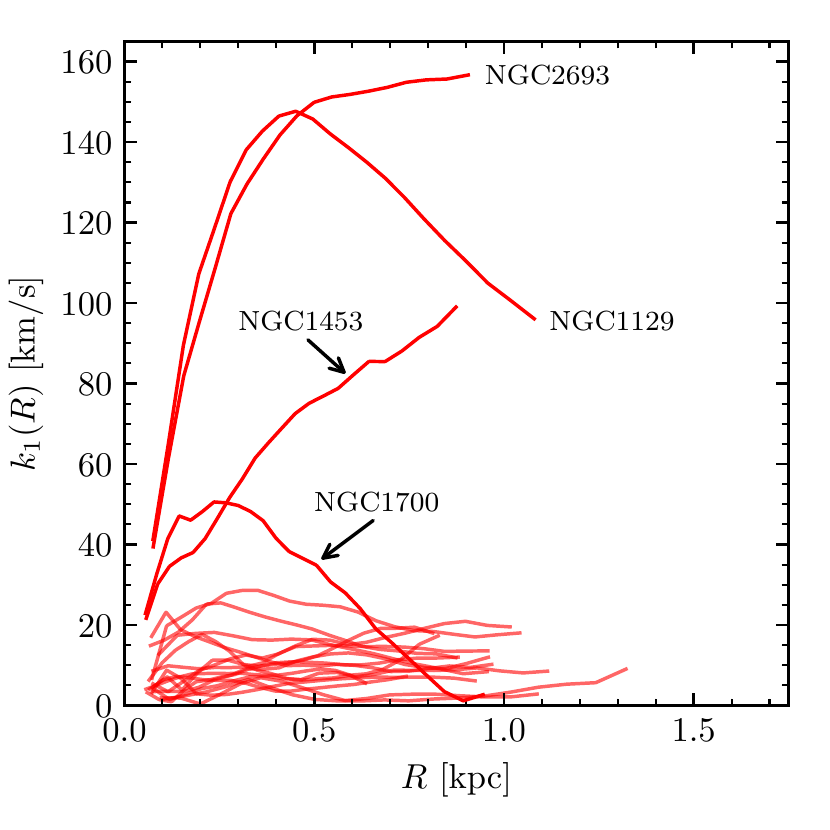}
    \caption{
    Velocity profiles within the central $\sim 1.5$ kpc for the 20 MASSIVE galaxies in this study, as measured by the the velocity extracted along the global kinematic axis PA$_{\rm kin}^{\rm GMOS}$ (top panel) and by the kinemetry coefficient $k_1$ (bottom panel).
    The four highlighted galaxies -- NGC~1129, NGC~1453, NGC~1700, and NGC~2693 -- show prominent central rotation, while the remaining 16 galaxies rotate slowly with $|V|, k_1 \lesssim 30 $ \kms.
    Overall, we find good agreement between the observed velocity field $V(R)$ and the kinemetry model velocity field $k_1(R)$ for all galaxies in our sample.
    }
    \label{fig:V_k1_profiles}
\end{figure}

The kinematic PA presented in Section~\ref{sec:central_rotation} quantifies the \emph{average} orientation of stellar rotation measured within an aperture.
The finely-resolved velocity maps from our GMOS IFS observations (Figure~\ref{fig:test}), however, often show intricate structures and contain more information than a single PA value.  In this section
we investigate these local features and analyze how the velocity (Section~\ref{sub:velocity_profiles})
and kinematic axis (Section~\ref{sub:kinematic_position_angle_profiles}) vary as a function of radius.

\subsection{Velocity profiles}
\label{sub:velocity_profiles}

The top panel of Figure~\ref{fig:V_k1_profiles} shows the velocity profiles along the kinematic axis measured from the GMOS data.
All but four galaxies rotate slowly (if at all) with $|V| \lesssim 30$ \kms, which is consistent with the results in Sec.~4.1.
For comparison, the radial profile of the $k_1$ coefficient defined in Eq.~(\ref{eqn:kinemetry}) from the kinemetry analysis is shown in the bottom panel of Figure~\ref{fig:V_k1_profiles}, where $k_1(R)$ traces the velocity along the radially changing kinematic position angle $\Gamma(R)$.
For galaxies with nearly constant $\Gamma(R)$ profiles, $|V(R)|$ and $k_1(R)$ are nearly indistinguishable (the four labeled profiles).
For the remaining galaxies with more complex velocity maps, there are subtle differences in the two profiles mainly because $k_1(R)$ follows the velocity along a kinematic axis that can twist significantly with radius.

\subsection{Kinematic position angle profiles}
\label{sub:kinematic_position_angle_profiles}

\begin{figure*}
    \includegraphics[width=\textwidth]{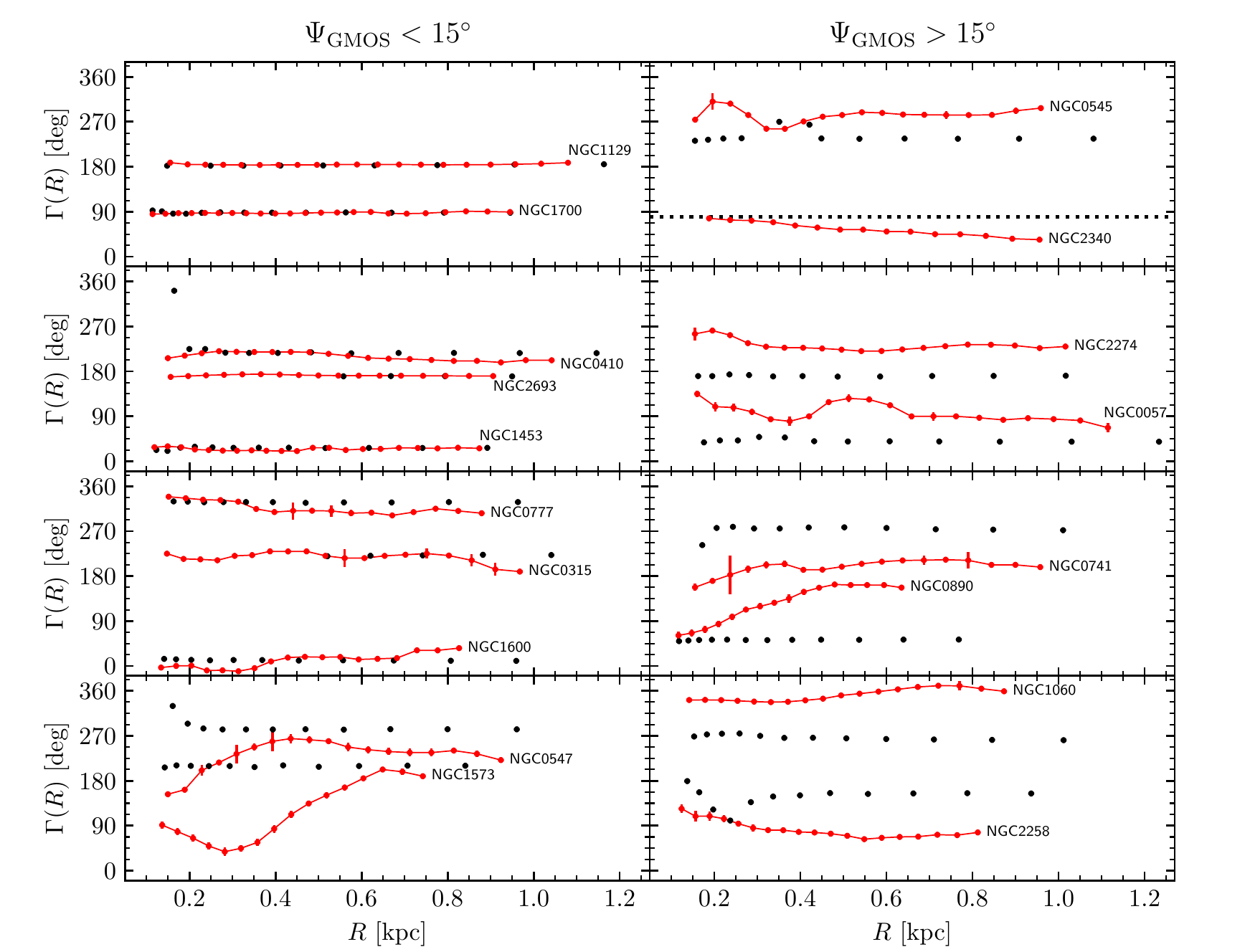}
    \caption{Radial profiles of the local kinematic position angle, $\Gamma$, measured across the GMOS 
    velocity maps for the 18 MASSIVE galaxies with detectable rotations (connected colored dots).  
    The local photometric PA measured from our HST WFC3 observations \citep{Goullaudetal2018} is overplotted (black points) for comparison. (The photometric PA is offset by a constant $180 \degr$ where needed to provide a closer match to $\Gamma(R)$.)
     The five galaxies in the two top-left panels all show very regular rotation in the central 1 kpc, with $\Gamma(R)$ changing by less than $20\degr$.
    The remaining 13 galaxies show kinematic twists of varying degrees.
    The galaxies in the left panels have small central misalignment angle (\psiG\ $< 15\degr$) and
    the ones in the right panels have \psiG\ $> 15\degr$.
   }
    \label{fig:gamma_profiles_all}
\end{figure*}

Figure~\ref{fig:gamma_profiles_all} shows the local kinematic PA angle, $\Gamma(R)$, from the kinemetry analysis (Section~\ref{sub:spatially_resolved_velocity_profiles}) for the 18 galaxies with measurable GMOS rotation. 
The local photometric PA (black dots) determined from our HST WFC3 observations is also shown.
Overall, we find that 5 out of 18 galaxies exhibit regular rotation across the GMOS FOV, while the remaining 13 show kinematic twists relative to the photometric axis to various extents.

The 10 galaxies classified as aligned rotators are shown in the left panels of Figure~\ref{fig:gamma_profiles_all}.
Four of the five galaxies shown in the two top panels (NGC~1129, NGC~1453, NGC~1700, and NGC~2693) exhibit very regular $\Gamma(R)$ profiles and tight alignment with the local photometry within the central $\sim 1$ kpc.
The fifth galaxy, NGC~410, also shows regular rotation with $\Gamma(R)$ varying by less than $\sim 20\degr$ and the departure from the photometric profile is small, but more noticeable than in the previous cases.
For the next three galaxies (NGC~315, NGC~777, NGC~1600), the kinematic axis twists by moderate amounts ($ < 40\degr$) around the photometric axis.
On average, however, the kinematic axis is still aligned with the photometric axis.

The bottom panel shows the two aligned galaxies (NGC~547 and NGC~1573) where PA$_{\rm kin}^{\rm GMOS}$ and PA$_{\rm phot}$ agree within the large errors on the kinematic axis.
Both galaxies show significant twisting in the kinematic axis of $\sim 100\degr$, and $\Gamma(R)$ is misaligned from the local photometric axis for most of the radial extent.
The amplitude of the detected rotation, however, is mostly below 10 \kms.   In the case of NGC~1573, we find a marginal drop in $k_{1}(R)$ from $\sim 7$ \kms\ to 4 \kms\ around 0.5 kpc where $\Gamma(R)$ shows strong twists. This is indicative of a KDC, but higher-resolution spectra would be needed to confirm it.

The remaining 8 galaxies have more complex GMOS velocity maps, where the kinematic axis twists with radius and the rotation is generally misaligned with the local photometric axis, as shown in the right panels of Figure~\ref{fig:gamma_profiles_all}.
The majority show moderate kinematic twists of $\sim 50\degr$, with only NGC~57 and NGC~890 showing extreme twists of $\sim 100\degr$ or larger.
As mentioned earlier, such large twists in $\Gamma(R)$ likely arise due to the very low velocities: $k_{1}(R)$ is less than 9 \kms~for both galaxies.

The detailed $\Gamma(R)$ profiles show that massive ETGs often exhibit complex features in their velocity maps.
These complex features are not fully captured by the simpler aligned/misaligned classification based on global kinematic properties (Section~\ref{sub:misalignment_between_kinematics_and_photometry}), as revealed by the fact that half of the aligned galaxies (left panel of Figure~\ref{fig:gamma_profiles_all}) actually show noticeable twists in the kinematic axis and deviations from the local photometric profile.


\section{Distinct Kinematic Features}
\label{sec:distinct_kinematic_features}

In this section we highlight six galaxies in the GMOS sample with distinct central kinematic features that indicate unusual assembly histories in their past. Four of them exhibit rotation around the photometric major axis rather than the typical minor axis.
Two galaxies have distinct central versus main-body kinematic and photometric features.
We discuss each of them here.

\subsection{Minor-axis rotations}
\label{sec:minor-axis}

As Figure~\ref{fig:test}(b) and Figure~\ref{fig:gamma_profiles_all} show,
four galaxies in our sample -- NGC~741, NGC~890, NGC~1060 and NGC~2258 -- have central kinematic axis that is nearly orthogonal to the photometric major axis (with $\Psi_{\rm GMOS} \ga 75 \degr$).
Furthermore, Figure~\ref{fig:pakin_scatter} shows that NGC~1060 and NGC~2258 have aligned inner and outer kinematic axes; these two galaxies as a whole are therefore rotating along their respective photometric minor axes. 
Such cases where the rotation is primarily \emph{around} the photometric major axis (equivalently \emph{along} the photometric minor axis) are sometimes said to be exhibiting ``minor-axis" or ``prolate-like" rotation (e.g. \citealt{SchechterGunn1979, DaviesBirkinshaw1986, DaviesBirkinshaw1988, Franxetal1989, JedrzejewskiSchechter1989}).

Several recent studies have found that a significant fraction of massive ETGs show minor-axis rotation.
In \cite{Eneetal2018} we find that 11 of 90 MASSIVE galaxies ($\sim 12$\%) with $M_* \gtrsim 10^{11.5} M_\odot$ exhibit minor-axis rotation with $\Psi > 60^\circ$ and 7 galaxies with $\Psi > 75^\circ$ on scales of $\sim 10$ kpc.
\cite{Tsatsietal2017} identified minor-axis rotation in 8 massive galaxies from the CALIFA survey \citep{Walcheretal2014}.
In their case, the minor-axis rotation occurs either in a kinematically distinct central component or in the galaxy as a whole.
They find that among massive ETGs, minor-axis rotation is present in $\sim 27$\% of CALIFA galaxies and $\sim 23$\% of ATLAS$^{\rm 3D}$ galaxies with $M_* \gtrsim 10^{11.3} M_\odot$.
For galaxies more massive than $10^{12} M_\odot$, \cite{Krajnovicetal2018} find that 44\% of their 25 MUSE galaxies show significant rotation around the photometric major axis.  A detailed analysis of 900 simulated ETGs in the Magneticum cosmological simulations (box size 68 Mpc; force softening $\ga 1$ kpc) finds about 20 galaxies in the mass range of the MASSIVE survey; among them, 3 are classified as prolate rotators (Fig.~B1 of \citealt{Schulzeetal2018}). Within the small number statistics, this result is in line with our survey result for main-body kinematics.  Simulations
with better force softening would be needed to study the finer kinematic features observed in our GMOS data.

\begin{figure}
    \includegraphics[width=\columnwidth]{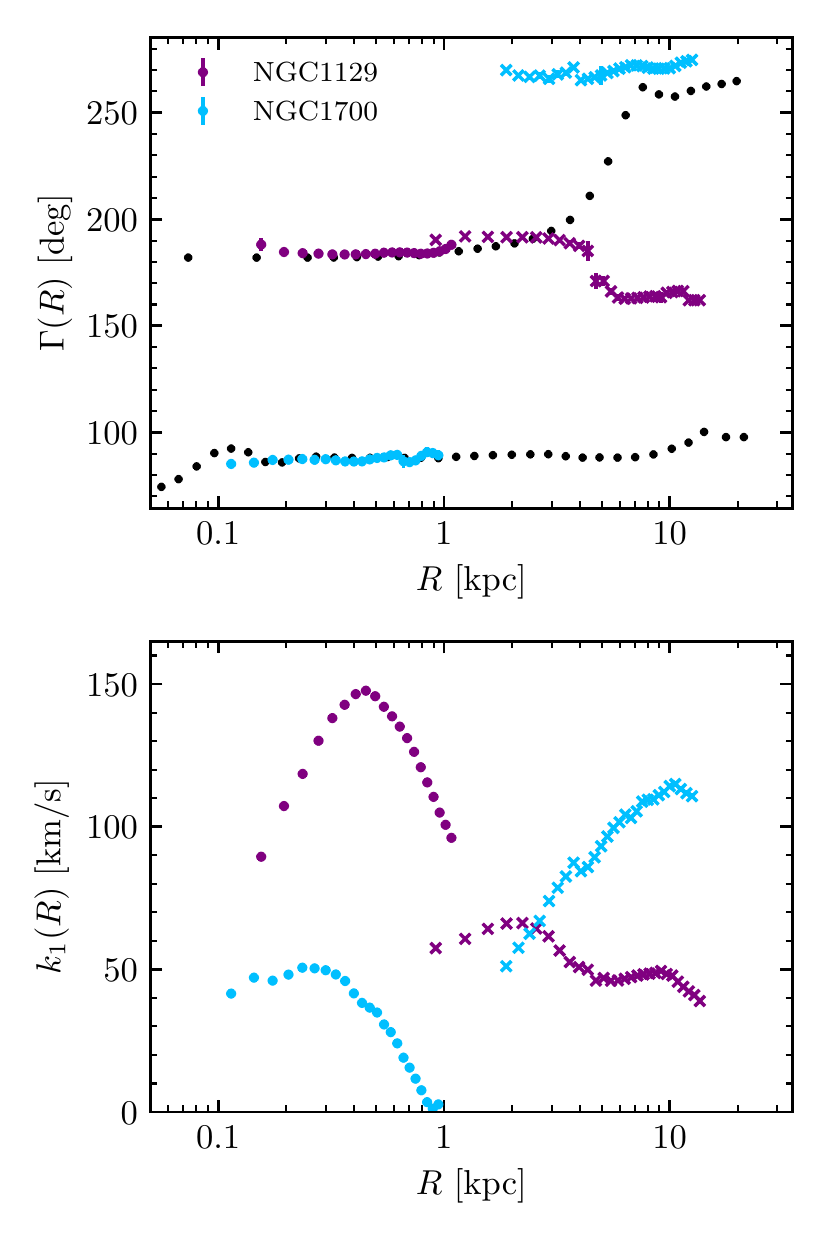}
    \caption{Radial profiles of the kinemetry position angle $\Gamma$ (top panel) and the kinemetry coefficient $k_1$ (bottom panel) across the GMOS (circles) and Mitchell (crosses) velocity maps for NGC~1129 and NGC~1700.
    The photometric PA radial profiles (offset by $180 \degr$ for NGC~1129) are shown with black points.
    For both galaxies, the local kinematic and photometric profiles agree within the central $\sim 1$ kpc, but show significant differences of $\sim 100 \degr$ and $\sim 180 \degr$, respectively, at larger radii.
    Additionally, the $k_1(R)$ profiles show a local peak within $R \sim 1$ kpc, suggesting the presence of distinct rotation components in the centers of these galaxies.
    }
    \label{fig:gamma_profiles_kdc}
\end{figure}

\subsection{NGC~1700: counter-rotating core}

As seen in Figures~\ref{fig:test}(a) and
\ref{fig:gamma_profiles_kdc} (top panel),
both the inner (blue dots) and outer (blue crosses) kinematic axes of NGC~1700 are well aligned with its photometric major axis (black dots). All three axes show little radial variations except for an abrupt $180\degr$ reversal in the rotational direction at $\sim 1$ kpc.  
The two distinct kinematic components are also clearly seen in the velocity amplitude profile, $k_1(R)$, in the bottom panel of Figure~\ref{fig:gamma_profiles_kdc}, 
where $k_1(R)$ reaches a local maximum of $\sim 50$ \kms~at $R \sim 0.3$ kpc before dropping to zero at $R \sim 1$ kpc.
It then smoothly increases to $\sim 120$ \kms\ at $R \sim 10$ kpc.

The counter-rotating core in NGC~1700 was not seen in our previous kinemetry analysis of the Mitchell observations \citep{Eneetal2018} because this core region ($\sim 1$ kpc or $\sim 4''$ in radius)
is below the resolution scale of the Mitchell IFS.
\citet{Franxetal1989} saw a hint of a distinct core in NGC~1700 as the innermost two velocity points in their long-slit data changed signs, but they cautioned that the results ``need confirmation." This confirmation is now provided by our finely-resolved velocity map for ~1700, which shows a striking and unambiguous counter-rotating core.

The distinct kinematic core of NGC~1700 
also has a distinct younger stellar population compared to the main body, suggesting that 
NGC~1700 is a product of a minor merger between the main galaxy and a small companion galaxy on a retrograde orbit \citep{Kleinebergetal2011}.
Our velocity dispersion map of NGC~1700 shows a single peak at the center (Fig.~23, \citealt{Eneetal2019}). It therefore is not a so-called 2$\sigma$ galaxy seen in a handful of lower-mass S0 galaxies (e.g., \citealt{Krajnovicetal2011}), which may have formed from a single major merger of two disk galaxies or via gas accretion (e.g., \citealt{Crockeretal2009, Boisetal2011, Katkovetal2016, Pizzellaetal2018}).

\subsection{NGC~1129}

For NGC~1129, Figure~\ref{fig:gamma_profiles_kdc} (top panel) shows that 
the local kinematic axis, $\Gamma(R)$, measured from our GMOS (magenta circles) and Mitchell (magenta crosses) data agrees well with each other in the inner $\sim 5$ kpc, and it shows little radial variations and is well aligned with the photometric major axis (upper black dots) in this region.
The accompanying $k_1(R)$ profile shows a strong velocity peak of $\sim 150$ \kms\ at $R \sim 0.5$ kpc (bottom panel of Figure~\ref{fig:gamma_profiles_kdc}).
Altogether, the inner few kpc of NGC~1129 resembles that of a typical fast regular rotator without any misalignment.  

Between $R\sim 3$ kpc and 8 kpc, however, the photometric PA of NGC~1129 shows a striking $\approx 90\degr$ twist, which was first reported in our HST-WFC3 study \citep{Goullaudetal2018}.
This transitional region was shown to be an inflection point in the ellipticity radial profile, which led us to suggest that NGC~1129 has recently undergone a major merger event.
Now our kinemetry results give further support to this claim, as the top panel of Figure~\ref{fig:gamma_profiles_kdc} shows that 
the kinematic axis $\Gamma(R)$ (magenta crosses) also changes at $\sim 5$ kpc, albeit with a smaller amplitude of $\sim 30\degr$.  


\section{Discussion}
\label{sec:discussion}

The diversity in the spatial variations of stellar velocity features and in the degree of misalignment between kinematics and photometry found in this study suggests diverse assembly histories for the  present-day massive ETGs.
Comparison with results predicted by numerical simulations often offer insight into the details of galaxy merger histories.

For kinematic features, galaxy merger simulations find that gas-rich major mergers of disc galaxies typically result in axisymmetric oblate elliptical galaxies with little kinematic misalignment, while gas-poor major mergers preferentially produce triaxial or prolate shaped elliptical galaxies that show a wide distribution in the kinematic misalignment angle (e.g., \citealt{NaabBurkert2003, Coxetal2006, Naabetal2014, Jesseitetal2009, Moodyetal2014, Yangetal2019}).
These predictions are in broad agreement with our observational results that the distributions of both the central and main-body kinematic misalignment angles peak at small values and extend all the way up to maximal misalignment.

The observed distributions of kinematic misalignment and ellipticity can be used to infer the distribution of the intrinsic shapes of the galaxies.  Our main-body data are found to be consistent with a population of mildly triaxial galaxies on average \citep{Eneetal2018}, but galaxy to galaxy variations are expected within the population of massive ETGs.
Furthermore, numerical simulations find that higher mass galaxies are more likely to be intrinsically prolate and that prolate galaxies often show minor-axis rotation \citep{EbrovaLokas2017, Lietal2018}.
We find 20\% of the sample galaxies to exhibit significant minor-axis rotation in the central $\sim 1$ kpc, and previously we find 11 of the 90 MASSIVE galaxies to show minor-axis rotation out to $\sim 1 R_e$ \citep{Eneetal2018}.
We note that while minor-axis rotation is consistent with a prolate shape, there is not a one-to-one correspondence between the two features. 
For instance, intrinsically prolate galaxies in the Illustris simulations are found to range from showing no rotation to
being kinematically aligned \citep{Lietal2018, BassettFoster2019}.

Simulations find that mergers impart a cumulative effect,
and galaxies that have experienced successive multiple mergers are more likely to have lower spins \citep{Lagosetal2018}. Furthermore,  slow-rotating remnants from multiple mergers of disc galaxies are more likely to show kinematics twists, while remnants from binary mergers most often do not \citep{Moodyetal2014}.  
The majority of galaxies in the MASSIVE sample
have low spins in both the central region and the main body,
and many of the low-spin galaxies in Figure~\ref{fig:gamma_profiles_all} exhibit noticeable kinematic variations.
These results are overall consistent with multiple gas-poor mergers as a main (albeit not only) formation pathway for massive ETGs.

A number of studies have investigated the origins of kinematic distinct components in ETGs.
One scenario is a minor merger between the main galaxy and a small companion galaxy on a retrograde orbit, which would lead to a remnant hosting a counter rotating core (e.g., \citealt{Kormendy1984, BalcellsQuinn1990}).
This is a plausible explanation for the central counter-rotating component in NGC~1700, since the central component was found to have a  distinct, younger stellar population compared to the main body \citep{Kleinebergetal2011}.
Other proposed explanations for kinematically distinct components include complex projection effects in the core of a triaxial system rather than a physically distinct entity \citep{Statler1991,vandenBoschetal2008}, a central component formed from the major merger of two disk galaxies \citep{Schweizeretal1990, HernquistBarnes1991, Hoffmanetal2010, Boisetal2011}, and a central disk formed from counterrotating accreting gas \citep{FranxIllingworth1988,Bertolaetal1998}.
The last formation scenario is likely the explanation for the small-scale ($\lesssim 0.3$ kpc) kinematical cores in fast-rotating SAURON galaxies \citep{McDermidetal2006}.


\section{Summary}
\label{sec:summary}

In this paper we have presented a detailed study of stellar velocity features 
in the central few kpc of 20 ETGs in the MASSIVE survey.
These galaxies are located at a median distance of $\sim 70$ Mpc and have stellar mass $M_* \gtrsim 10^{11.7} M_\odot$.
The finely-resolved velocity map for each galaxy is obtained from high-$S/N$ ($\sim 120$) spectra from the Gemini GMOS IFS with a $0.2''$ spatial sampling and $5''\times 7''$ FOV, covering a physical length scale of $\sim 100$ pc to $\sim 2$ kpc.
This is a fraction of the typical effective radius of  $\sim 10$ kpc for MASSIVE galaxies (column 4 of Table~1; \citealt{Vealeetal2017b, Goullaudetal2018}). 
Prior characterizations of the spins and kinematic misalignments of ETGs have largely been based on measurements of stellar velocities over the main-body of the galaxies at scales of $\sim 1 R_e$.

Combining these central kinematics with our wide-field ($107''\times 107''$)  
main-body kinematics of MASSIVE galaxies \citep{Eneetal2018}, we have analyzed the stellar velocity profiles and the relative alignments of the central kinematic axis, main-body kinematic axis and the photometric major axis.

Our main findings are:

\begin{itemize}
\item 16 of the 20 galaxies have low spins ($\lambda \la 0.1$) and low rotation velocities (below 50 \kms) in both the central region and the main-body.  Our earlier findings that massive ETGs with $M_* \gtrsim 10^{11.5} M_\odot$ are predominantly slow rotators \citep{Vealeetal2017a, Vealeetal2017b} therefore also apply to the central kpc of these galaxies.  
   
\item  Only 4 galaxies have high $k_1^{\rm max} > 50$ \kms\ in both central and outer parts, with NGC~1453 and NGC~2693 showing similar $k_1^{\rm max}$ values, and NGC~1129 and NGC~1700 showing very different $k_1^{\rm max}$ values at the center versus at $1 R_e$.
   
\item We measured the kinematic position angle PA$_{\rm kin}^{\rm GMOS}$ for 18 of the 20 galaxies; the remaining two galaxies (NGC~1016 and NGC~4874) have no detectable central rotations.
We found 10 of the 18 galaxies to have aligned central kinematic and photometric axes with small misalignment angle ($\Psi_{\rm GMOS} \lesssim 15\degr$).  
For the remaining 8 galaxies, $\Psi_{\rm GMOS}$
is distributed quite evenly from $15\degr$ to the maximum value of $90\degr$, where
four galaxies (NGC~741, NGC~890, NGC~1060 and  NGC~2258) exhibit ``minor-axis" or ``prolate-like" rotation with $\Psi_{\rm GMOS} \ga 75 \degr$.
This distribution of the central kinematic misalignment is very similar to that of the main-body misalignment angle for 71 MASSIVE galaxies presented in \cite{Eneetal2018}.
    
\item 
We found a strong correlation between central kinematic misalignment and galaxy spin, again similar to our earlier main-body result \citep{Eneetal2018}.
The clear trend is that $\sim 90$\% of galaxies with high spins ($\lambda \ga 0.2$) have well aligned kinematic and photometric axes, while only $\sim 40$\%
of low-spin galaxies are well aligned.

\item Despite the similarities between the central and main-body rotation statistics above, the two kinematic axes within individual galaxies are not always aligned.  Only 12 galaxies in our joint datasets exhibit sufficient rotations for us to determine both kinematic axes, but we observed a diverse range of alignment configurations even within this small sample.
Only 7 of the 12 galaxies have aligned central and main-body kinematic axes. Among them, 
the two kinematic axes are also aligned with the photometric axis in 5 galaxies, whereas the two kinematic axes in the other 2 galaxies (NGC~1060 and NGC~2258) are almost perpendicular to the photometric axis and hence exhibit ``minor-axis" rotations.

For the 5 galaxies with misaligned central and main-body kinematic axes, we observed three types: (1) central kinematic axis aligned with photometric axis but not with main-body kinematic axis (NGC~410 and NGC~777); (2)
the two kinematic axes and the photometric axis are all different from one another (NGC~890 and NGC~2258);
and (3) a counter-rotating inner core that is anti-aligned with the main-body rotation by $180^\circ$ (NGC~1700).
    
\item To make further use of the GMOS velocity maps
beyond measuring the averaged central spin and kinematic axis,
we analyzed the radial profile of the {\it local} kinemetry position angle $\Gamma(R)$, which traces the direction of rotation at a given radius. 
We found 13 galaxies to exhibit kinematic twists of $\ga 20 \degr$ in the central $\sim 2$ kpc.
The kinematic twists are not limited to galaxies with large central misalignment angle.
A handful galaxies with $\Psi_{\rm GMOS} \lesssim 15\degr$ show noticeable local kinematic twists.

\end{itemize}

We have found that the central kpc regions of massive ETGs exhibit diverse velocity features that range from regular rotations to kinematically distinct components.
 These detailed features could be uncovered only with high-$S/N$ and high-resolution spectroscopic and photometric observations that span two orders of magnitude in radial coverage.
The diversity of the observed kinematic features suggests that local massive ETGs have complex merger histories.
Cosmological numerical simulations that properly capture the large-scale galaxy environments as well as resolve sub-kiloparsec scale kinematics at redshift 0 are needed for a full assessment of the formation pathways of massive ellipticals and for statistical comparisons with current observational results.
The intricate velocity structures shown in this work further suggest that stellar orbit libraries containing all allowed orbital classes would be needed to fully sample the phase space
of these massive ETGs and to perform dynamical mass modeling of the central supermassive black holes in these galaxies.

\acknowledgments

We thank the anonymous referee for a careful reading of the manuscript and useful suggestions. The MASSIVE survey is supported in part by NSF AST-1411945, NSF AST-1411642, NSF AST-1815417, NSF AST-1817100, HST GO-14210, HST GO-15265 and HST AR-14573.
CPM acknowledges support from the Heising-Simons Foundation, the Miller Institute for Basic Research in Science, and the Aspen Center for Physics, which is supported by NSF grant PHY-1607611. J.L.W. is supported in part by NSF grant AST-1814799.
This work is based on observations obtained at the Gemini Observatory, processed using the Gemini IRAF package, which is operated by the Association of Universities for Research in Astronomy, Inc., under a cooperative agreement with the NSF on behalf of the Gemini partnership: the National Science Foundation (United States), National Research Council (Canada), CONICYT (Chile), Ministerio de Ciencia,
Tecnología e Innovación Productiva (Argentina), Ministério da Ciência, Tecnologia e Inovação (Brazil), and Korea Astronomy and Space Science Institute (Republic of Korea).

\end{document}